# The equilibrium theory of life evolution

D. Balciunas

**Abstract**. Apparent biodiversity on earth exists only if we compare different species separated from their environments. Meanwhile coexisting species have to be identical in terms of energetic interactions. Consider the biosphere as a network of chemical reactions. This leads to the conclusion that forces which drive life evolution may be found inside the process of the transformations of molecules. From the thermodynamic point of view the system of reacting particles reaches a steady state when the chemical potentials of reagents become equal. Interactions between chemical compounds taking part in concurrent reactions and the equilibration of their chemical potentials is the essence of biological evolution and its only driving force.

Today scientists believe that species in ecosystems evolve through a sophisticated mechanism hypothesized by Darwin and named natural selection (*e. g.* Schneider and Helms, 2003; Biebricher and Eigen, 2005; Lieberman *et al.*, 2005). Natural selection is thought to occur at molecular level as well (Eigen and Shuster, 1979; William *et al.*, 2005). However such a mechanism - natural or artificial - is an illusion. Darwin based his theory on anthropomorphic arguments such as favorable variations, struggle for existence, selection (Darwin, 1859). Thus it reflects only human emotional response to processes occurring in the living world. Darwinian mechanism of evolution - natural selection idea - is just a new kind of the phlogiston theory.

The first steps in the formation of living things on earth were chemical evolution (Des Murais and Walter, 1999). So we should look for biological evolution driving forces in the dynamics of chemical reactions instead of searching for any particular biological principles.

At first I briefly describe the mathematical expressions I will use further. Suppose two self-replicating components with a total constant mass density $M$ make a system. Components exchange their masses with the environment and interact with each other by transferring mass from one component to another. Then the dynamics of a particular component $x$ mass density may be described by the rate equation

$$v = \pm\, \rho\, x\, (1 - x\, /\, K) \tag{1}$$

Here $v$ is the growth rate of component $x$ mass density. $\rho$ denotes the maximal instantaneous growth rate for a mass density unit corresponding to given parameters, i.e. $v/x \rightarrow \rho$ when $x \rightarrow 0$ and $y \rightarrow M$ ($y \neq x$). Function $K(\rho) \equiv M - c_y$ is the instantaneous stable mass density of component $x$, where $c_y$ means an instantaneous equilibrium point of component $y \neq x$ at the same time. I assume that always $K > 0$. Which sign to write in (1) depends on the component selected. In the case one of the components has different forms $x_1, x_2, ..., x_n$ a set of equations will be used

$$v_i = \pm\, \rho_i\, x_i\, (1 - x_1\, /\, K_i - x_2\, /\, K_i - ... - x_n\, /\, K_i) \tag{2}$$

In this expression $K_i(\rho_i) \equiv M - c_{y(i)}$. System (2) describes the evolution of competing components $x_i$ in the most general form.

Let's start with a formal reversible autocatalytic chemical process

$$s + p \rightarrow p + p \qquad\qquad (3)$$
$$p \rightarrow s$$

(3) may be the simplest model of life on earth. In this case one should bear in mind that the backward reaction is driven by the external force, e.g. by the energy obtained from external sources, e.g. from sunlight or the planet itself. On the other hand system (3) may present only a part of an ecosystem food chain. Then the backward process $p \rightarrow s$ means that the total mass density of the system remains constant.

According to (1) the evolution of system (3) particle density $x$ is described by the equation

$$x(t) = x(t_0) \pm \left( \int_{t_0}^{t} \rho x \, d\tau - \int_{t_0}^{t} (\rho x^2 / K) \, d\tau \right) \qquad\qquad (4)$$

Here $x$ represents the particle density of either $s$ or $p$. Sign "–" corresponds to the model written for $s$ component. $K \equiv M - c_y$ and $c_y$ are as described above. The second term in the parentheses describes virtual interactions between different particles of a component. From this model we obtain that if the components of the system coexist permanently mean values of $x$ and $y$ satisfy the following condition $x \rightarrow c_x$, $y \rightarrow c_y$ when $t \rightarrow \infty$.

$$\lim_{t \rightarrow \infty} t^{-1} \int_{t_0}^{t} (u - c_u) \, d\tau = 0, \quad u \in \{x, y\} \qquad\qquad (5)$$

Alternatively at equilibrium the dissipative function for a unit of system (3) volume may be presented in the form

$$- dG / dt = \mu_s v^+ - \mu_p v^- \qquad\qquad (6)$$

where $G$ is the Gibbs free energy. $v^+$ and $v^-$ are the rates of the forward and back reactions respectively. Chemical potentials $\mu_i$'s describe "forces" acting in the system at equilibrium. When the system is approaching equilibrium both $v^+ - v^- \rightarrow 0$ and $\mu_s - \mu_p \rightarrow 0$.

Now let's incorporate competitors in our system. If two or more concurrent reactions, involving divergent competitors $p$ or convergent competitors $s$, proceed at the same time the evolution equation for a particular component $x_i$ particle density following (2) will be

$$x_i(t) = x_i(t_0) \pm \left( \int_{t_0}^{t} \rho_i x_i \, d\tau - \int_{t_0}^{t} (\rho_i x_i x_1 / K_i) \, d\tau - \dots - \int_{t_0}^{t} (\rho_i x_i x_n / K_i) \, d\tau \right) \qquad (7)$$

To analyze coexistence in the system it is worth to split the virtual competition terms in equation (7) into two parts, symmetric and anti-symmetric. Then we obtain the following expression

$$x_i(t) = x_i(t_0) \pm \left( \int_{t_0}^{t} \rho_i x_i \, d\tau - \Sigma \int_{t_0}^{t} (\rho_i x_i x_k / K_k) \, d\tau - \dots - \int_{t_0}^{t} \varphi_{ik} x_i x_k \, d\tau \right) \qquad (8)$$

Only anti-symmetric terms are responsible for the coexistence of different component forms. Functions $\varphi_{ik} = (c_{y(i)} - c_{y(k)}) / K_i K_k$ ($i = 1, \dots, n$; $k = 1, \dots, n$) express the relative effect of directed competition in the system. If competitors coexist permanently, then

$$\lim_{t \to \infty} t^{-1} \int_{t_0}^{t} \varphi_{ik} \, x_i x_k \, d\tau = 0 \qquad (9)$$

From here we obtain that at equilibrium

$$\lim_{t \to \infty} t^{-1} \int_{t_0}^{t} (c_{y(i)} - c_{y(k)}) \, d\tau = 0 \qquad (10)$$

This is in accordance with expression (6) for concurrent reactions. If the system is in equilibrium the chemical potentials of all competing forms $x_i$ must be equal.

Transferring the above reasoning to ecosystems we may write the rate equations for a particular set of competing species $x_i$ mass densities dynamics

$$v_i = f_i - g_i \qquad (11)$$

Here $f_i$'s are functions describing the rate of mass gain per unit of volume from a common source if the equations are written for d-competitors; or $g_i$'s are functions which mean the rate of mass loss per unit of volume due to grazing effect by a common predator (for c-competitors). Then we may write equations analogous to (8) for every competing component mass density dynamics. In this case $c_{y(i)} = f_i \, y \, / \, g_i$ for c-competitors or $c_{y(i)} = g_i \, y \, / \, f_i$ for d-competitors.

Symmetric parts describe relations between species as if they were identical, what means they equally use energy. Anti-symmetric parts show differences in competitors energy allocation. I named these two aspects of interaction process indifferent and directed competition respectively (Balciunas, 2004). Species which equally use energy are identical. Nature cannot distinguish between them. The overall conclusion is that only identical species may coexist – the result which has been missed by 20th century theoretical ecology. That happened due to spread of wrong basic ideas such as the logistic and the Lotka-Volterra competition equations (Lotka, 1925), the competitive exclusion principle (Hardin, 1960), the limiting similarity principle (MacArthur and Levins, 1967).

Biological evolution does not mean that species become more adapted to their environment due to the process of natural selection. From the physical point of view life on the planet is a network of chemical reactions. Evolution only means that all (electro)chemical potentials in the biosphere tend to become equal. It is no more purposeful and no more "intelligent" than a chemical reaction in a test tube. No matter what striking biosphere looks for us, it is simply the result of a grand chemical experiment on concurrent reactions.